\documentclass[12pt,preprint]{aastex}

\slugcomment{Draft version: \today}

\begin{document}

\title{A Structural Analysis of Star-Forming Region AFGL~490}
\author{L.~C.~Masiunas\altaffilmark{1}, R.~A.~Gutermuth\altaffilmark{2,3}, J.~L.~Pipher\altaffilmark{3,4}, S.~T.~Megeath\altaffilmark{5}, P.~C.~Myers\altaffilmark{6}, L.~E.~Allen\altaffilmark{7}, H.~M.~Kirk\altaffilmark{6}, G.~G.~Fazio\altaffilmark{6}}
\altaffiltext{1}{Five College Astronomy Dept., Smith College, Northampton, MA}
\altaffiltext{2}{Dept. of Astronomy, University of Massachusetts, Amherst, MA}
\altaffiltext{3}{Visiting Astronomer, Kitt Peak National Observatory, National Optical Astronomy Observatory, which is operated by the Association of Universities for Research in Astronomy (AURA) under cooperative agreement with the National Science Foundation.}
\altaffiltext{4}{Dept. of Physics \& Astronomy, University of Rochester, Rochester, NY}
\altaffiltext{5}{Dept. of Physics \& Astronomy, University of Toledo, Toledo, OH}
\altaffiltext{6}{Harvard-Smithsonian Center for Astrophysics, Cambridge, MA}
\altaffiltext{7}{National Optical Astronomy Observatory, Tucson, AZ}

\begin{abstract}
We present {\it Spitzer} IRAC and MIPS observations of the star-forming region containing intermediate-mass young stellar object (YSO) AFGL~490.  We supplement these data with near-IR 2MASS photometry and with deep SQIID obervations off the central high extinction region.  We have more than doubled the known membership of this region to 57 Class~I and 303 Class~II YSOs via the combined 1-24~$\mu$m photometric catalog derived from these data.  We construct and analyze the minimum spanning tree of their projected positions, isolating one locally over-dense cluster core containing 219 YSOs (60.8\% of the region's members).  We find this cluster core to be larger yet less dense than similarly analyzed clusters.  Although the structure of this cluster core appears irregular, we demonstrate that the parsec-scale surface densities of both YSOs and gas are correlated with a power law slope of 2.8, as found for other similarly analyzed nearby molecular clouds.  We also explore the mass segregation implications of AFGL~490's offset from the center of its core, finding that it has no apparent preferential central position relative to the low-mass members.
\end{abstract}

\section{INTRODUCTION\label{intro}}

The {\it Spitzer Space Telescope} has provided a tremendous leap in sensitivity and spatial resolution over previous infrared telescopes, enabling numerous surveys of nearby star-forming regions \citep[e.g.][]{meg04,tei06,mue07,gut09} as well as unbiased surveys of entire star-forming molecular clouds \citep[e.g.][]{jor06,har06,eva09,reb10,reb11}, revealing most of the members that have dusty circumstellar material in these regions.  Such an unbiased view has demonstrated that despite the prevalence of the clustered formation environment \citep{lad03,por03}, less than a quarter of these stars are found in regions dense enough to experience close approaches with neighbors that would have notable effects on circumstellar disks \citep{bre10}.  The high frequency of clustered star formation gives insight into the physical processes that govern where and when molecular clouds form stars \citep{hei10,gut11}.  Therefore, in order to understand the process by which clouds give rise to young stars, we must compare the structure of star-forming regions to that of their natal clouds.

The many {\it Spitzer} surveys have brought an empirical basis against which we can compare any new data from an individual star-forming region.  For example, \citet{eva09} found 1024 YSOs across five nearby star-forming clouds and improved the constraints on star formation rates and efficiencies at low and high densities.  \citet{reb11} studied the nearby North America Nebula, expanding the known population of the region by an order of magnitude.  \citet{gut09} found 2548 YSOs in 36 nearby star-forming regions, providing a comprehensive analysis of the clusters' structures and forming a broad picture of star formation within the nearest kpc.

Within this context, we have performed a focused investigation of the AFGL~490 star-forming region.  The bright infrared source AFGL~490 (also known as GL 490), which lies at a distance of 900~pc \citep{sne84} in the Cam OB1 association, has been well-studied at many wavelengths from near-IR to millimeter.  It resides $\sim$5~pc away from the variable star CE Cam, an exposed A0I star whose UV emission is likely responsible for excitation of the strong 8.0~$\mu$m emission from polycyclic aromatic hydrocarbon (PAH) dust in the region.  AFGL~490 is a 8-10~M$_{\sun}$ protostar that is variable in the near-infrared \citep{jon90}; it has long been known to have a bipolar molecular jet \citep{sne84}, an extended envelope, and a rotating disk \citep{sch06}.  Another large outflow in the region, AFGL~490-iki, was discovered $7^{\prime}$ north of AFGL~490 \citep{pur95}.  One of the first mentions of a cluster associated with AFGL~490 is found in a paper by \citet{hod94}, which reported that the cluster was small (45 associated stars) and had a low density.  Most recently, the AFGL~490 region was studied by \citet{gut09}, more than tripling the membership of the region; however, the field of view of the reported observations truncated the apparent distribution of the cluster.

Here we present an expanded infrared survey of the AFGL~490 region, including new {\it Spitzer} and deep ground-based near-IR imaging, with the aim of fully sampling its spatial extent.  The paper is organized as follows.  In Section~\ref{data}, we describe the data, its reduction, and source catalogues.  In Section~\ref{analysis}, we describe our source classification and the products of our data, including extinction and n$^{th}$ nearest neighbor maps.  In Section~\ref{discuss}, we analyze the data products in the context of structure: we find the densest area in the region and compare it to a previous survey, and investigate the possible mass segregation of AFGL~490.  In Section~\ref{summary}, we summarize our results.

\section{DATA \& OBSERVATIONS\label{data}}

\subsection{IRAC Imaging\label{irac}}
\begin{figure}
\epsscale{.80}
\plotone{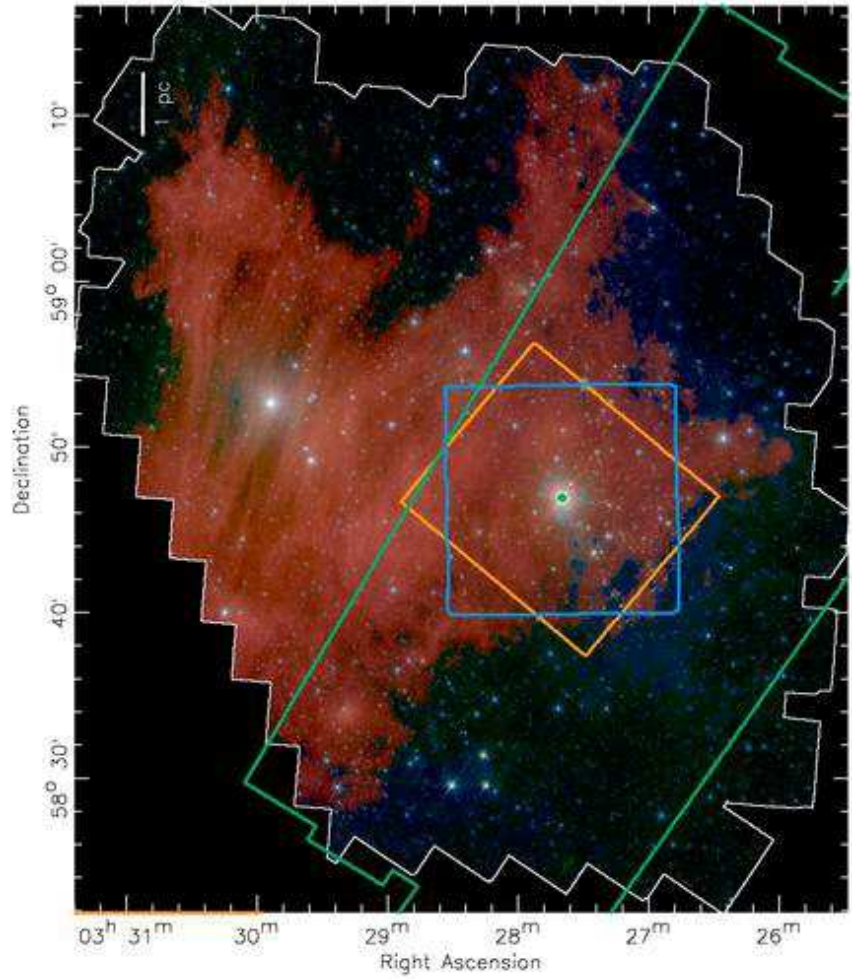}
\caption{Coverage of IRAC (background image, outlines in white), overlaid with the
  original IRAC coverage in \citet{gut09} (orange), SQIID JHK coverage
  (blue), and expanded MIPS 24~$\mu$m coverage (green).  For the color image,
  blue is 3.6~$\mu$m, green is 4.5~$\mu$m, and red is 8.0~$\mu$m.
\label{coverage}}
\end{figure}

AFGL~490 was initially observed using {\it Spitzer Space Telescope}'s IRAC instrument at 3.6, 4.5, 5.8, and 8.0~$\mu$m on 2004 October 8 (AOR ID 3654656, centered on Right Ascension $03^{h}27^{m}38.60^{s}$ and Declination +$58^{\circ}46^{\prime\prime}57.7^{\prime}$, J2000) as part of the Guaranteed Time Observation Proposal ID 6, ``Structure and Incidence of Young Embedded Clusters.''  These data were analyzed and presented by \citet{gut09}, where it was noticed that the YSO distribution of the region extended to the edge of the surveyed $15^{\prime}$x$15^{\prime}$ field of view (orange overlay in Figure~\ref{coverage}).  The IRAC view of the region was expanded to 0.51 square degrees as part of the GTO ``Clusters Near Clusters'' program (PID 40147).  These data were obtained on 2007 October 23 (AOR ID 21894144) and 2008 March 9 (AOR ID 21894400).  Integration time was 10.4~sec~pix$^{-1}$ with four dithered images, giving an effective integration time of 41.6 seconds.  All observations were obtained in High Dynamic Range mode, where  10.4 second and 0.4 second observations were taken at each pointing to extend the effective dynamic range of the survey.

The data were initially processed by the S18.7.0 software package at the Spitzer Science Center, producing standard basic calibrated data (BCD) products.  Post-BCD data treatment was performed using Cluster Grinder (CG), a {\it Spitzer} data analysis package written in IDL \citep{gut09}.  In summary, bright source artifacts were removed, correcting ``banding'' and ``pulldown,'' as well as cosmic ray hits and other transient artifacts.  Mosaics were made at a pixel resolution of 0.86$^{\prime\prime}$~pix$^{-1}$, with an effective resolution of 2.2$^{\prime\prime}$ at FWHM; a color image of the IRAC mosaics is found in Figure~\ref{coverage}.  CG incorporates PhotVis version 1.10 \citep{gut08} to automatically detect point sources using an algorithm that robustly isolates stars in fields of varying background emission.  Source positions measured from the IRAC data were compared with 2MASS counterpart coordinates to improve the astrometric calibration of each BCD image before making final IRAC mosaics.  Aperture photometry was performed using an aperture radius of $2.4^{\prime\prime}$, with background inner and outer annuli of $2.4^{\prime\prime}$ and $7.2^{\prime\prime}$ respectively.  Photometric zero-points adopted (ie. the magnitude assigned for 1~DN~s$^{-1}$ flux in an aperture) were 20.207, 19.452, 17.251, and 17.644 magnitudes for 3.6, 4.5, 5.8, and 8.0~$\mu$m bands, respectively.  These zero points are derived from standard calibration values and aperture corrections\citep{rea05}, and account for our reduced pixel size.  The field-averaged 90\% differential completeness limits are 16.6, 16.4, 14.8, 13.3 magnitudes in the same bands, respectively, with a signal to noise requirement of 5 or greater (0.2 magnitudes).  These are derived from artifical star insertion and retreival testing following the method of \citet{gut05}.

\subsection{MIPS Imaging\label{mips}}
Initial 24~$\mu$m MIPS images for PID 6 \citep{gut09} were obtained on 2004 September 24 (AOR ID 3663104, centered on Right Ascension $03^{h}28^{m}31.6^{s}$ and Declination +$58^{\circ}25^{\prime\prime}37^{\prime}$, J2000.0).  The field of view was expanded on 2007 October 23 (AOR ID 21894144) as part of PID 40147.  MIPS data covers 0.38 square degrees (green overlay in Figure~\ref{coverage}), 0.26 of which are within IRAC coverage.  All observations were taken at the medium scan rate with an effective integration time of 40.4 seconds.

The data were initially processed by version S16.1.0 of the standard MIPS BCD pipeline.  Post-BCD processing was performed with Cluster Grinder (described in Section~\ref{irac}).  Mosaics were made at a pixel resolution of 1.8$^{\prime\prime}$~pix$^{-1}$ with an effective resolution of 4.5$^{\prime\prime}$ at FWHM.  Aperture photometry was performed using an aperture radius of $7.6^{\prime\prime}$ with background inner and outer annuli of $7.6^{\prime\prime}$ and $17.8^{\prime\prime}$ respectively.  The photometric zero-point adopted was 15.35 magnitudes \citep[14.6 from][with conversion for smaller pixel size]{gut08}, and the field averaged 90\% differential completeness limit is 8.4 magnitudes with a signal to noise requirement of 5 or greater (0.2 magnitudes).

\subsection{SQIID Imaging\label{sqiid}}
\begin{figure}
\epsscale{.80}
\plotone{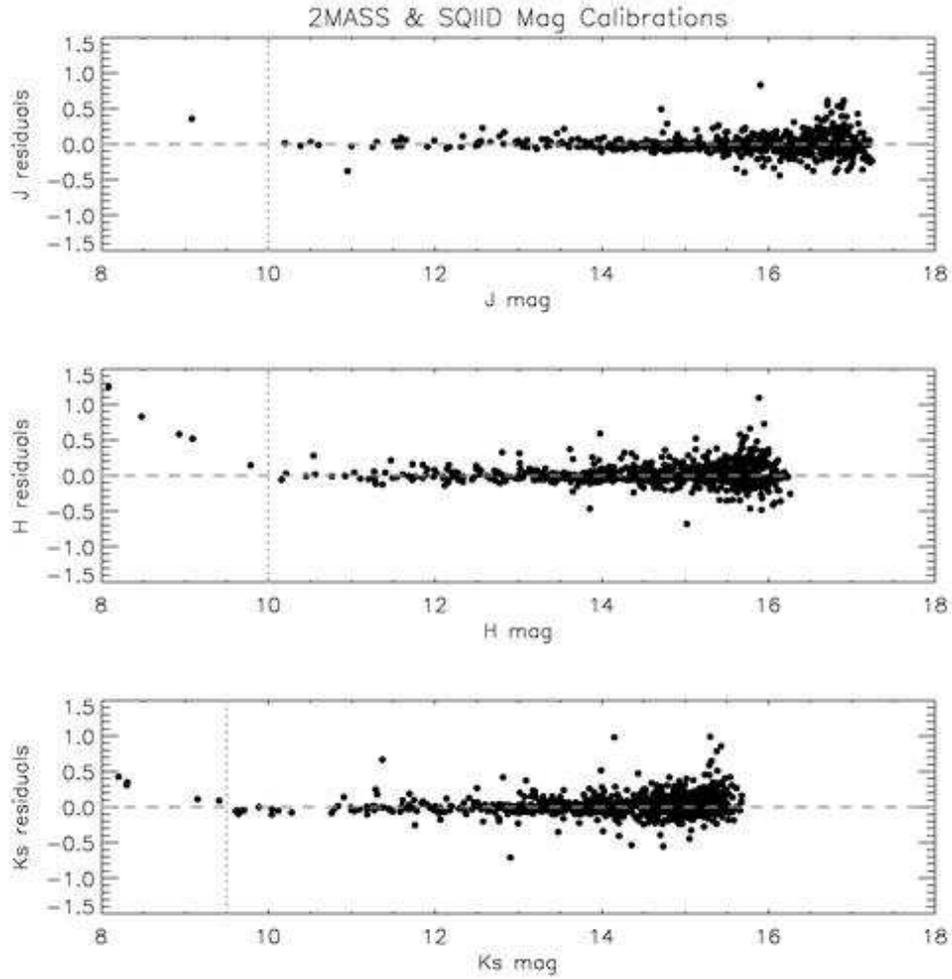}
\caption{Plot of the magnitude residuals from the merged SQIID and
  2MASS catalogue.  The x-axis plots the 2MASS magnitudes, while the
  y-axis plots the difference between 2MASS and SQIID magnitudes.  Dotted lines
  mark the saturation point for SQIID at J=10, H=10, and K$_{S}$=9.5 magnitudes.\label{mag_resids}} 
\end{figure}

Near-infrared imaging in J, H, and K$_{S}$ bands were obtained using the Simultaneous Quad Infrared Imaging Device (SQIID) on Kitt Peak National Observatory's 2.1-m telescope on 2003 November 8.  Four one-minute dithers were taken per position in a 3x3 position raster.  The final mosaics have a mean integration time per position of 240 seconds, and cover 0.05 square degrees ($14^{\prime}\times14^{\prime}$, blue overlay in Figure~\ref{coverage}), with a platescale of $0.70^{\prime\prime}$~pix$^{-1}$ and a FWHM of 2.6, 2.5, and 2.6 pixels for J, H, and K$_{S}$ bands, respectively.  SQIID data were reduced using custom IDL routines for ground-based near-IR data reduction \citep{gut05}.  Those routines include modules for linearization, flat-field creation and application, background frame creation and subtraction, distortion measurement and correction, astrometric calibration, cosmic ray filtering, and mosaicking. 2MASS data were used as the astrometric reference for distortion measurement and final position calibration.  A source catalogue was made using PhotVis version 1.10, with radii of the apertures and inner and outer limits of the sky annuli of $2.1^{\prime\prime}$, $3.5^{\prime\prime}$, and $5.5^{\prime\prime}$, respectively, and a sigma threshold of 9.  Photometric zero-points were 22.43, 22.42, and 21.83 magnitudes for J, H, and K$_{S}$ bands, adopted in order to minimize residuals with 2MASS photometry (Figure~\ref{mag_resids}).

\begin{figure}
\epsscale{.80}
\plotone{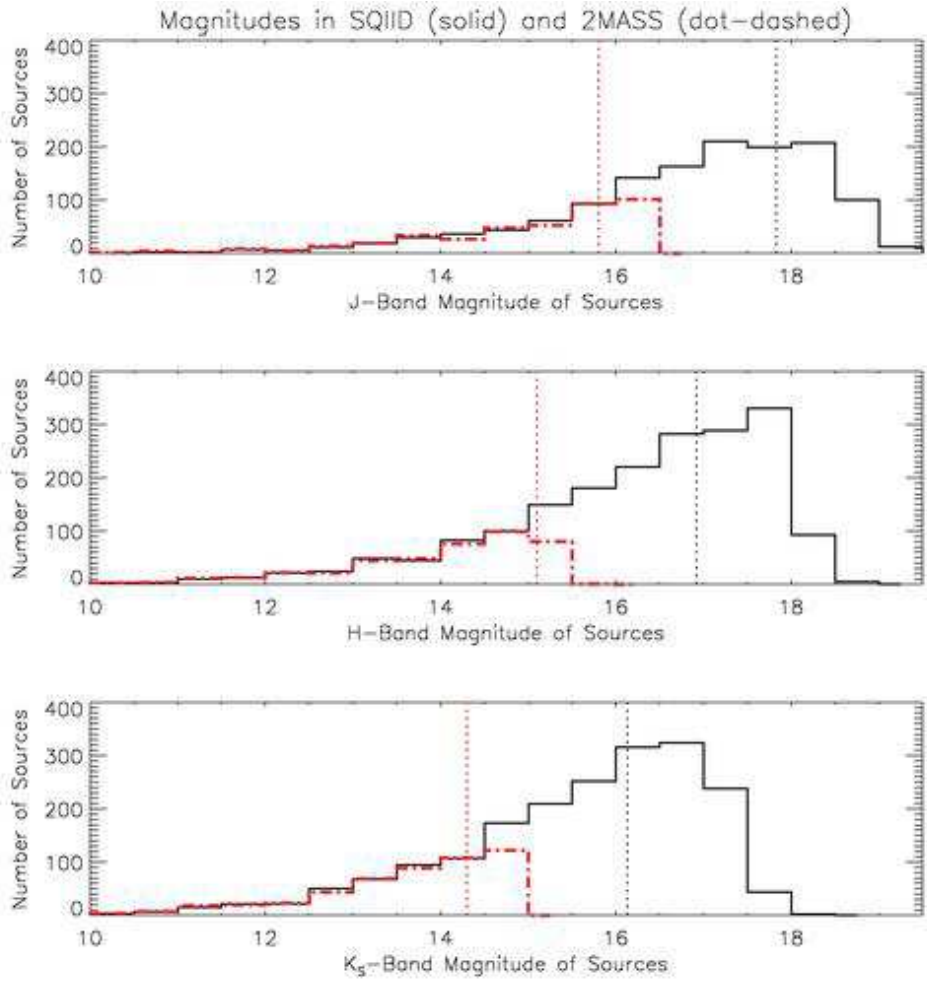}
\caption{SQIID (solid) and 2MASS (dot-dashed) source counts at
  different magnitudes, by wavelength, for the SQIID-observed region
  of the overall survey.  Vertical lines show completeness for each
  data set, with SQIID showing 90\% and 2MASS showing
  99\% completeness.\label{datasets}} 
\end{figure}

The SQIID data greatly increased near-IR sensitivity relative to 2MASS coverage in the densest area of the region.  These data are considerably deeper than 2MASS, reaching field-averaged 90\% differential completeness limits of 17.8, 16.9, and 16.1 magnitudes in J, H, and K$_{S}$ bands, respectively, with a signal to noise requirement of 10 or greater (0.1 magnitudes).  For comparison, 2MASS is 99\% complete at 15.8, 15.1, and 14.3 magnitudes in J, H, and K$_{S}$ bands.  A comparison of the SQIID and 2MASS data can be found in Figure~\ref{datasets}.  The median residuals for bright objects are 0.035, 0.040, and 0.045 magnitudes for $J$, $H$, and $K_S$, respectively; the dim source residuals are dominated by the poor signal to noise of the 2MASS detections: 0.08, 0.11, 0.11 magnitudes.

\subsection{Source Catalogue\label{cats}}
The final merged source catalogue includes data in J, H, and K$_{S}$ bands from 2MASS and SQIID; 3.6, 4.5, 5.8, and 8.0~$\mu$m from IRAC; and 24~$\mu$m from MIPS.  Where 2MASS and SQIID data overlap, the SQIID data are used, provided detections are dimmer than saturation limits of J=10, H=10, and K$_{S}$=9.5 magnitudes and have a good signal to noise ratio (here, a signal to noise ratio of 5).  If a detection in any SQIID band did not meet this criteria, that detection was replaced by the 2MASS data in each band in an effort for each source to have more consistent data.  To make the final catalogue, the IRAC and 2MASS source lists were merged with a limiting tolerance of $1^{\prime\prime}$, then MIPS detections were included where IRAC or 2MASS detections existed using a tolerance of $2^{\prime\prime}$.  Finally, the SQIID data were merged, using a tolerance of $1^{\prime\prime}$.  

\section{ANALYSIS\label{analysis}}

\subsection{Source Classification\label{class}}
We applied the three-phase 1-24~$\mu$m point source classification scheme of \citet{gut09} to our merged catalog.  In each phase of the technique, we use a unique subset of the available bandpasses to securely identify YSOs while minimizing contamination by field stars, extragalactic sources, and local nebulosity.  By mitigating bias from reddening by dust associated with the natal cloud material, we can confidently classify those YSOs as envelope-bearing protostars (called Class~I below, but includes some Class~0 and flat spectrum YSOs) or pre-main-sequence stars with circumstellar disks (called Class~II below, but includes transition disks).  Tests performed on this classification scheme suggest that the likelihood of confusing a Class~II with an edge-on disk as a Class~I are less than 3.3\% \citep{gut09}.  We do not classify diskless YSOs (Class~III) because they are indistinguishable from field stars; we cannot reliably identify them using this data set.

\begin{figure}
\epsscale{.80}
\plotone{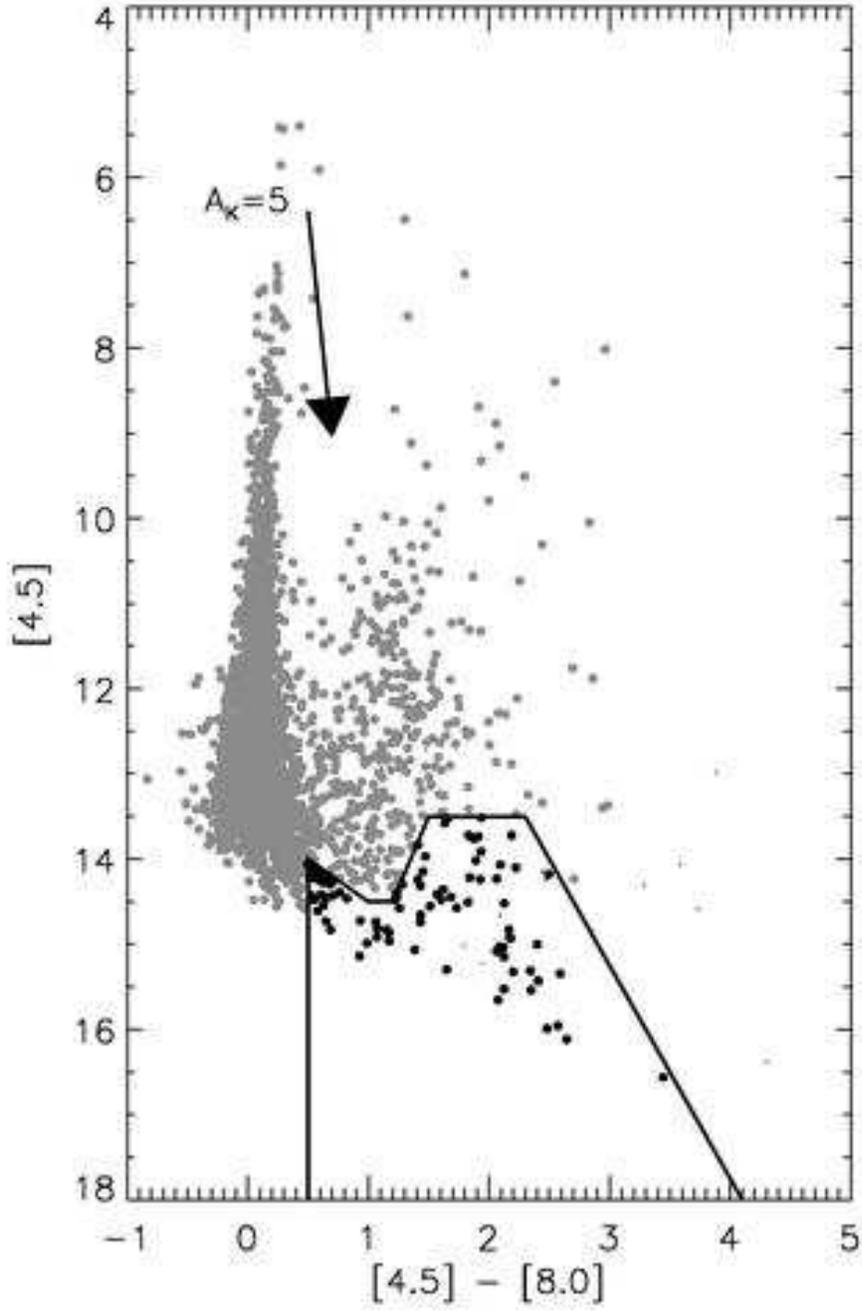}
\caption{Color-color diagram for the data used here, plotting [4.5] vs. [4.5]-[8.0], used in identifying AGN.\label{cm_224}} 
\end{figure}

\begin{figure}
\epsscale{.80}
\plotone{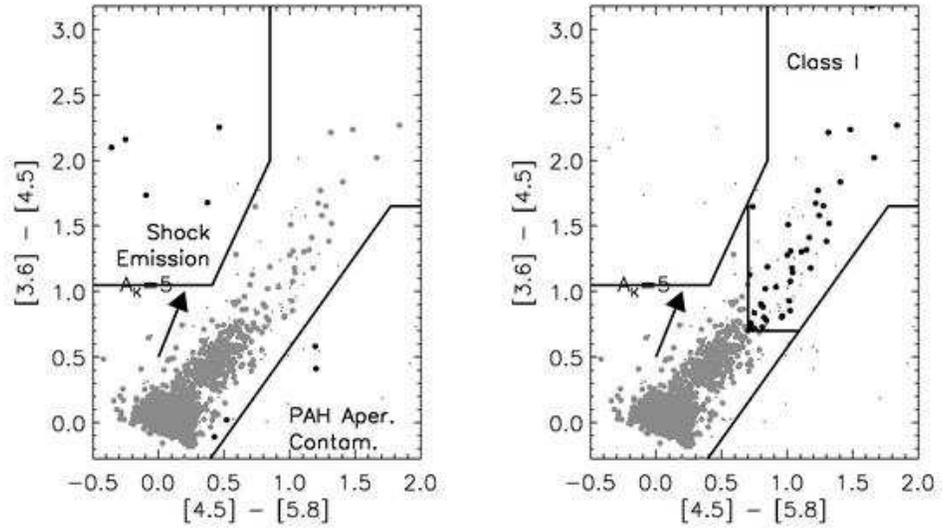}
\caption{Color-color diagram for the data used
  here, plotting [3.6]-[4.5] vs. [4.5]-[5.8], used in isolating
  unresolved shock emission knots, sources with structured PAH emission contamination, and Class I YSOs.\label{cc_1223}} 
\end{figure}

\begin{figure}
\epsscale{.80}
\plotone{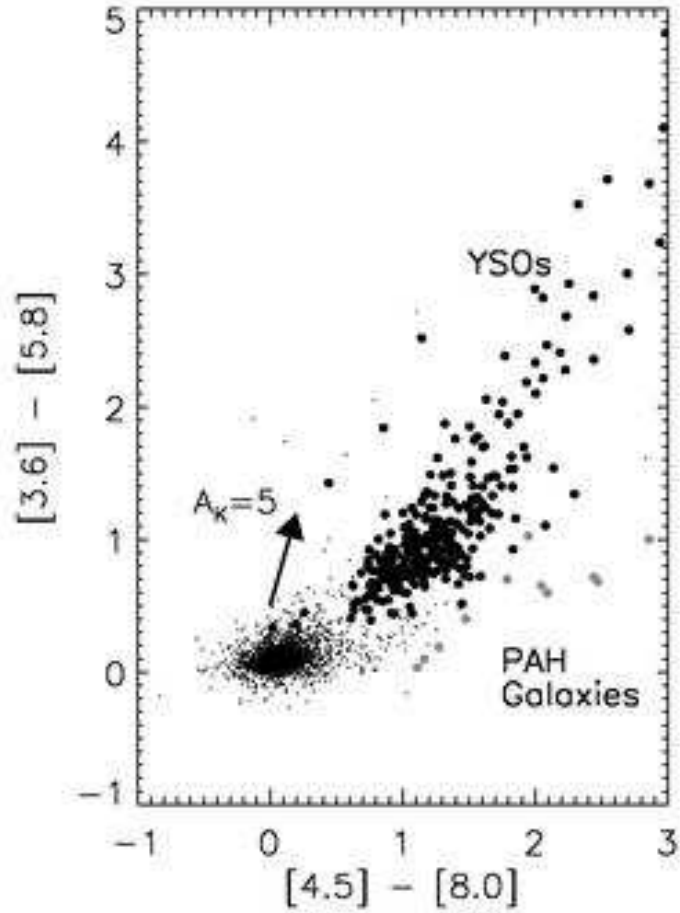}
\caption{Color-color diagram for the data used
  here, plotting [3.6]-[5.8] vs. [4.5]-[8.0], used in identifying both
  PAH galaxies and Class II sources (plotted here with Class I YSOs included).\label{cc_1324}} 
\end{figure}

In Phase 1 of this technique, we utilized only those sources detected in all of the {\it Spitzer} IRAC bands with photometric uncertainties of less than 0.2 magnitude.  Unresolved star-forming galaxies have a very distinct 5.8-8.0~$\mu$m color caused by strong PAH feature emission; we used custom color-color diagram cuts to uniquely identify and remove these sources.  Here, we also identified active galactic nuclei, which look like low luminosity YSOs with flat spectral energy distributions but follow unique conditions in 4.5 and 8.0~$\mu$m color space; they can be identified using cuts on a color-magnitude diagram (Figure~\ref{cm_224}).  Unfortunately, this method does not find all AGN: on average, it is estimated that 7 AGN per square degree are misidentified as YSOs \citep{gut09}, and some bona fide YSOs which are older or more distant are lost.  In this phase, we also removed unresolved knots of shock emission, which are bright at 4.5~$\mu$m because of strong molecular hydrogen emission.  Finally, we classified YSOs as Class~I or II predominantly with the [4.5]-[5.8] discriminant color (Figure~\ref{cc_1223}), which is affected much less by dust reddening than [3.6]-[4.5] \citep{fla07}.  A color-color diagram showing the basic resuts from this process is found in Figure~\ref{cc_1324}.

We applied Phase 2 to those sources which lack detections in 5.8 and 8.0~$\mu$m bands, a common occurrence among YSOs near the nebulous centers of embedded clusters, and which have high signal to noise detections in H and K$_{S}$ bands (uncertainty less than 0.1 magnitude).  First, we measured the photometric reddening of each source by dust along the line of sight as a ratio of $E_{H-K_{S}}$ using the source's photometry and the reddening law of \citet{fla07}.  The primary means of taking this measurement was the J-H vs. H-K$_{S}$ color-color diagram with the \citet{mey97} Classical T Tauri Star (CTTS) locus defining the intrinsic colors; if the source was not detected at J band, the H-K$_{S}$ vs. [3.6]-[4.5] color-color diagram with the YSO locus of \citet{gut05}, calibrated against the JHK CTTS locus, was used instead.  We then converted the $E_{H-K_{S}}$ color excess to $E_{K-[3.6]}$ and $E_{[3.6]-[4.5]}$ using the reddening law of \citet{fla07}, facilitating dereddening of those colors for use in YSO identification and classification via the K$_{S}$-[3.6] vs. [3.6]-[4.5] color-color diagram.  This allowed us to determine YSO classifications based on dereddened colors.  Also, we set a brightness limit in this phase requiring all Class~II YSOs to have [3.6]$_{0}~<$~14.5~magnitudes and protostars to have [3.6]$_{0}~<$~15~magnitudes, which allowed for another pass at filtering out dim extragalactic contaminants.

In Phase 3, we re-examined the entire catalogue, re-evaluating sources with a 24~$\mu$m detection with an uncertainty of less than 0.2 magnitude.  Objects classified as lacking in IR-excess in the earlier phases but having excess emission at 24~$\mu$m were classified as transition disk YSOs.  Those sources which had insufficient detections in J, H, K$_{S}$, and the IRAC bands but with bright 24~$\mu$m emission were considered deeply embedded objects (likely protostars).  Those sources which lacked a strong 24~$\mu$m detection but had been classified as Class~I YSOs in Phase 1 were reclassified as heavily reddened Class~II YSOs.

In total, we have identified 360 YSOs with infrared excess emission in AFGL~490.  Of those, 57 have excess consistent with Class~I YSOs and 303 have excess consistent with Class~II YSOs.  There are 22 sources classified as YSOs with transition disks (with large inner holes) and 10 as deeply embedded YSOs.  For the purposes of the following analysis, the deeply embedded protostars were included in the Class~I count as envelope-bearing protostars and the transition disk YSOs were included in the Class~II count as pre-main sequence stars with disks.  A complete list of the YSOs identified here can be found in Table~\ref{sourcelist}.

The YSO completeness is nontrivial to characterize, as physical parameters like luminosity and mass are rendered uncertain by poor constraints on heliocentric distance, age, and line of sight extinction by dust.  By adopting a distance of 900~pc, a pre-main sequence stellar age of 1~Myr, and the model color-magnitude tracks of \citet{bar98}, we find that we are complete to {\it diskless} 0.2~$M_{\odot}$ stars.  However, we have not identified any diskless YSOs as members, thus this represents an overestimate of the limiting mass for disk-bearing Class~II YSOs.  Under the assumption of a standard initial stellar mass function, we have detected over 90\% of the stars and over 99\% of the total stellar mass, modulated by the fraction of the membership that has a disk \citep{kro01}.  That fraction is expected to be greater than half.  Completeness for Class~I sources is more difficult to constrain based on a physical quantity, as the near- and mid-IR emission from these objects is dominated by the variable accretion luminosity which is reprocessed by the non-uniform, dusty outer envelope \citep{dun10}.

\subsection{n$^{th}$ Nearest Neighbor YSO Surface Density Map\label{nnd}}

\begin{figure}
\epsscale{.80}
\plotone{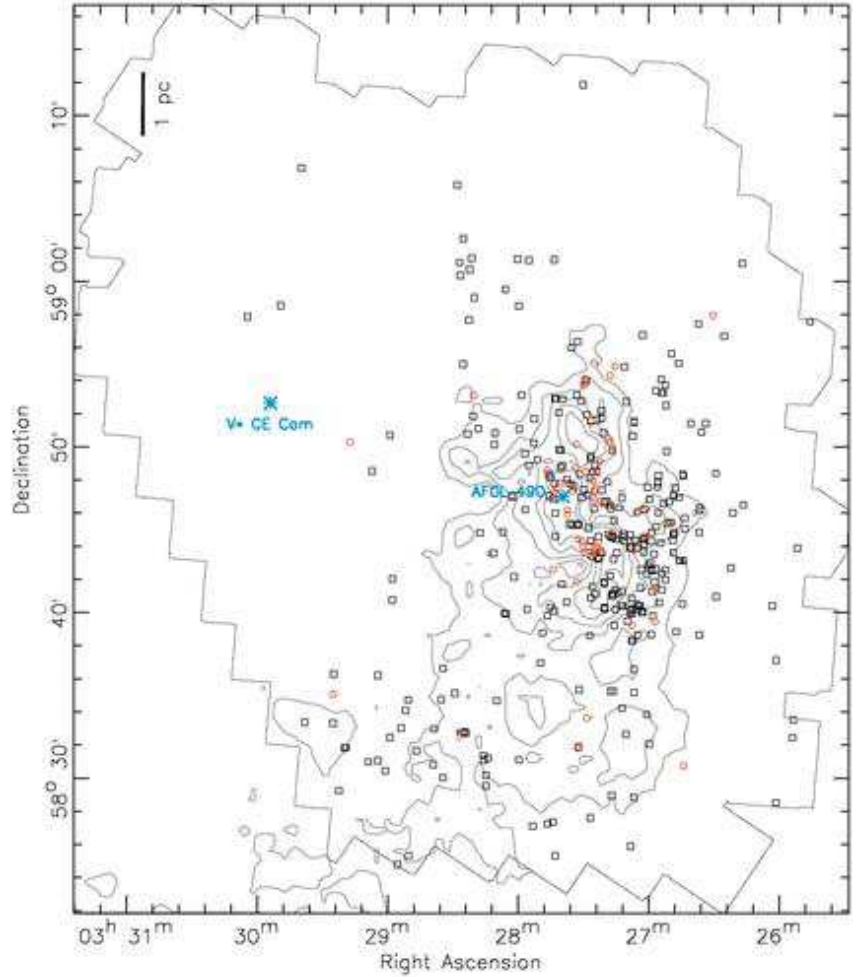}
\caption{Contours of near-IR extinction from dust, beginning with  A$_{V}$~=~5
  and increasing at intervals of 2 magnitudes.  The positions of YSOs
  are overlaid, with Class I and Class II sources represented in black and red,
  respectively.  The intermediate-mass star AFGL 490, embedded within the
  cloud, and variable star CE Cam, which is exposed, are plotted in blue.  Dark grey
  line delineates IRAC field of view.\label{av_map}}
\end{figure}

\begin{figure}
\epsscale{.80}
\plotone{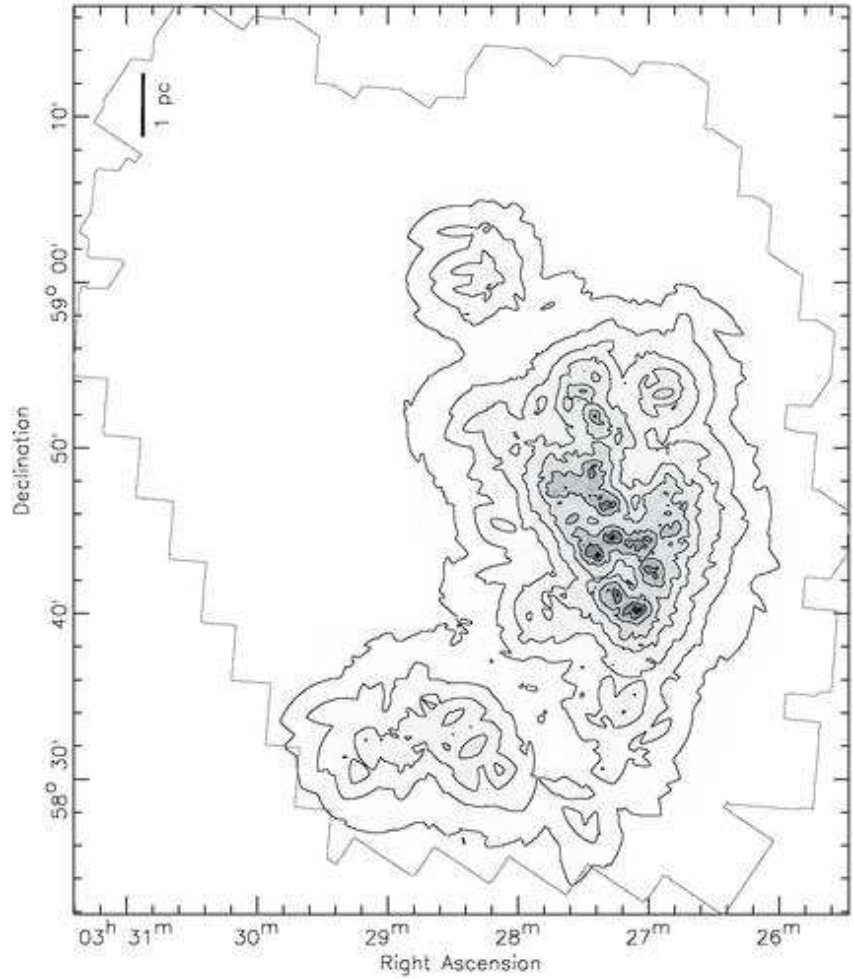}
\caption{Surface density map of YSOs, as determined from adaptively smoothing over the nearest six neighbors.  Greyscale is chosen such that 0 YSOs per square parsec is white and 200 is black.  Contours are overplotted, increasing at intervals of 1$\sigma$ ($\sigma$~=~50\% in the case of $n=6$) from the next highest level \citep{cas85}.  Dark grey line delineates IRAC field of view.\label{nn_map}}
\end{figure}

We present the spatial distribution of YSOs in Figure~\ref{av_map}, overlaid on the contours of near-IR extinction (described below).  The YSOs are visually overdense near AFGL~490, as well as north and southwest of it, apparently correlated with the regions where extinction is greater than 7 magnitudes.  To more easily compare the surface density structure of the YSOs to the extinction map, we made a smoothed map of their surface density.  To achieve this, we adopted the $n^{th}$ nearest neighbor surface density estimator smoothed over $n~=~6$ neighbors \citep{gut08}.  More specifically, at each location in a uniform grid, we calculated the radial distance to the sixth nearest YSO, $r_6$, and then computed the local surface density following the standard formula: $N(n)~=~(n-1)/(\pi~r_n^2)$ \citep{cas85}.  We chose a grid size of 9$^{\prime\prime}$ to Nyquist sample the shortest $r_6$ distance centered on each YSO.  The typical (median) $r_6$ is 70$^{\prime\prime}$; given that we used an adaptive smoothing technique, the effective resolution varies at any given point.  The range of surface densities in the map are 0.03 to 250~YSOs~pc$^{-2}$.  Figure~\ref{nn_map} shows a contour plot of the final surface density map, with contours increasing at intervals of 1$\sigma$ \citep[$\sigma$~=~50\% in the case of n~=~6,][]{cas85} from the next highest level overplotted.

\subsection{Near-Infrared Extinction Map\label{avmap}} 

In order to characterize the structure of molecular gas in the region, we constructed a map of near-IR extinction from dust using the combined SQIID and 2MASS catalogue.  We achieved this by computing the mean $H-K_S$ color excess of background stars (foreground stars are filtered via iterative outlier rejection) relative to an assumed intrinsic color $H-K_S~=~0.2$ for the nearest 20 sources to each point in a uniformly sampled grid \citep{gut05}.  As with the YSO surface density maps generated above, we adopt a grid size of 37$^{\prime\prime}$ for the extinction map to Nyquist sample the median $r_{20}$ distance of low extinction, and thus high density, field stars.  Because of the diminishing brightness of stars at higher extinctions, the adaptively smoothed map loses some natural resolution in those places \citep[r$_{20}$ ranges from 33$^{\prime\prime}$ to 129$^{\prime\prime}$][]{gut11}.  The final extinction map has a range of 0 to 19.8~$A_V$ (magnitudes).  The incorporation of the significantly deeper SQIID data into the near-IR catalog resulted in a much higher fraction of YSOs detected at H-band, causing spurious extinction structure in the map where high IR-excess sources are found.  To improve the quality of the map, we removed all {\it Spitzer}-identified YSOs from the catalog used to make the extinction map (Figure~\ref{av_map}).  Upon visual comparison of the extinction map, which shows gas structure, and YSO surface density map (Figure~\ref{nn_map}), which shows the structure in the YSO spatial distribution, the similarities are striking; we will investigate this similarity in Section~\ref{cloud}.

\subsection{Cluster Core Extraction and Structure\label{coreext}}

Figures~\ref{av_map} and \ref{nn_map} reveal that several locally overdense groupings, or {\it cluster cores}, are apparent in the smoothly varying spatial distribution of YSOs in AFGL~490.  We isolate these cluster cores following the methodology of \citet{gut09}, summarized below.

We began by connecting all of the YSOs to their nearest neighbors through a unique, unified network of lines such that the total connection length is minimized, called a minimum spanning tree \citep[MST,][]{car04}.  Every point was connected by the minimum distance between the next point to form a ``branch,'' similar to the way the eye connects points in a constellation \citep{kir11}.  It is worth noting that the nearest neighbor distances analyzed in Section~\ref{nnd} above are in fact a degenerate subset of the MST branches \citep{gut09}.  In other words, a sizable fraction of MST branches connect nearest neighbors and are double counted in the nearest neighbor distance distribution above.

\begin{figure}
\epsscale{.80}
\plotone{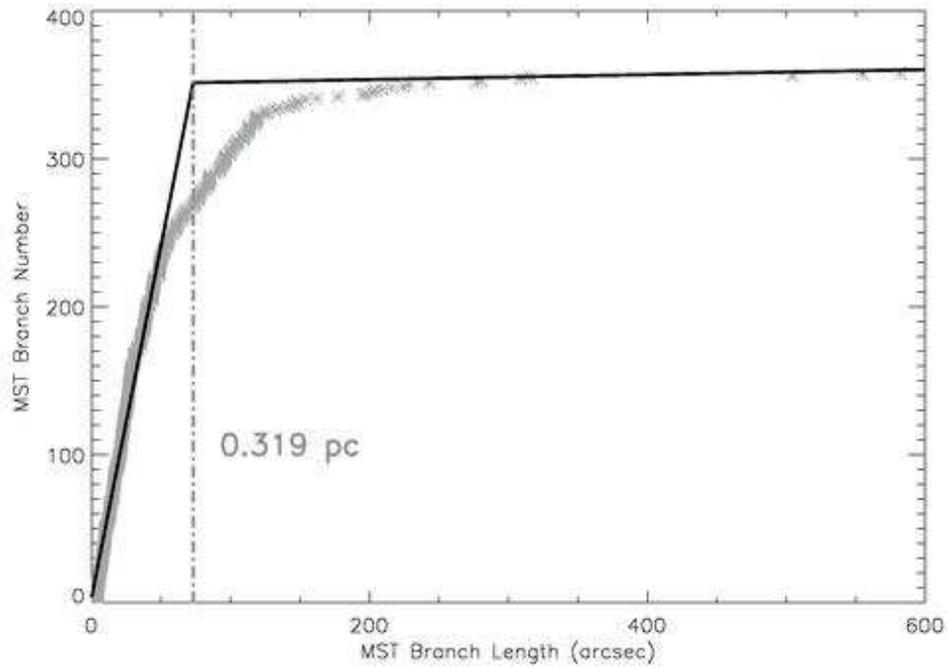}
\caption{Cumulative distribution of MST branch lengths for the YSOs in
  the region (asterisks).  The solid line segments are an approximation
  of the CDF; the steeply-sloped line corresponds to the denser
  region, while the shallowly-sloped line corresponds to the more
  isolated region.  The convergence of these two lines corresponds to
  a length which becomes the threshold branch length $L_{crit} $ (dot-dashed line).\label{mst_brkpt}}
\end{figure}

Once the MST was constructed, we defined cluster cores by adopting a threshold branch length, $L_{crit}$.  Any connections longer than this length, we erased.  To avoid choosing an arbitrary length to define a core \citep{schm06}, we used the algorithm defined in \citet{gut09} to determine the threshold length from the cumulative distribution of branch lengths.  First, we plotted the cumulative distribution function (CDF) of the MST branch lengths.  The CDF can be approximated by two intersecting line segments; one is steep and the other shallow, corresponding to the denser and more isolated areas in the region respectively.  We assign $L_{crit}$ as the branch length of the point where these two lines intersect.  The CDF, overplotted with the line segments and threshold branch length, can be found in Figure~\ref{mst_brkpt}.  We adopted this technique because it robustly identifies {\it relative} overdensities among the YSO distribution.  In essence, we have traded away the systematic grouping ambiguities that accompany the application of strict, uniform density thresholds \citep[e.g.][]{eva09} caused by incomplete membership counts (uncertain disk fraction) and heliocentric distances, in favor of a relative threshold definition that references the lowest density YSOs observed, and is thus prone to uncertainty from a limited field of view \citep{gut09}.  The expanded field of view of AFGL~490 presented here should address that uncertainty.  We also adopted a minimum membership threshold, $N_{min}$, among the isolated groups to reduce the likelihood of statistical fluctuations inducing false, small-N groups.  Therefore, we define a cluster core as a population of stars numbering at or greater than the minimum membership limit connected together by MST branch lengths shorter than the threshold branch length as measured above.

We find $L_{crit}~=~0.319$~pc with the above technique; under this threshold, we identify only a single, large core with membership greater than the adopted $N_{min}~=~10$ \citep{gut09}.  This core contains 219 YSO members: 47 Class~I protostars and 172 Class~II pre-main sequence stars with disks.  The core makes up 60.8\% of the region's membership.  It contains 82.5\% of the region's Class~I population and 56.8\% of the Class~II population.  Because of the relatively different durations of the Class~II and Class~I evolutionary phases, the ratio of the number of Class~II members to Class~I members is a proxy for the relative age of similarly observed regions.  The AFGL~490 cluster core has a Class~II/Class~I ratio of 3.7 (smaller than the regional Class~II/Class~I ratio of 5.3).  We draw a convex hull, a polygon whose verticies are the outermost members of a group of points, around the cluster core members to define the cores area and size.  Because of the nature of the convex hull, it was not capable of tracing the unique shape of the core; visually, the hull encompasses four extra sources that are not included in the cluster core.

\begin{figure}
\epsscale{.80}
\plotone{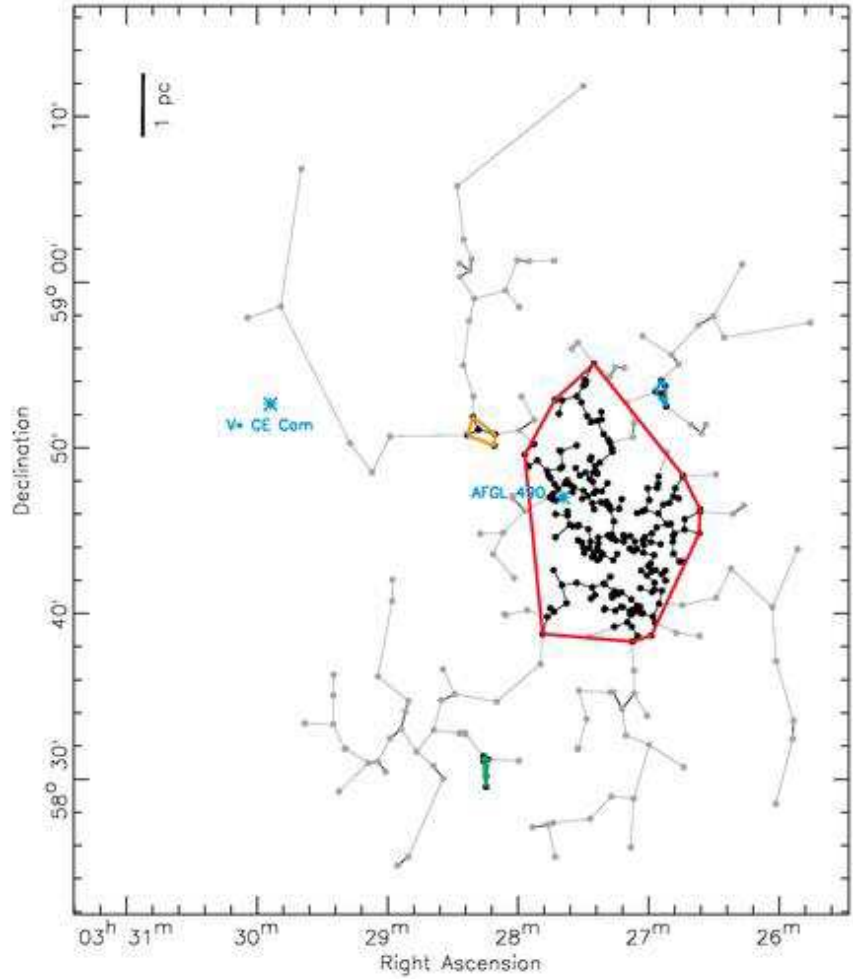}
\caption{The complete MST, with connections less than the threshold
  length $L_{crit}$ in black.  Cluster cores are overlaid in color.  The large
  cluster core in the middle includes 219 YSOs, and is the only core to be
  large enough to be considered according to the method followed in
  \citet{gut09}.  In the surrounding cores, one has six and the others
  have five members.\label{mst_grps}}
\end{figure}

It is interesting to note that when $N_{min}$ was reduced to five sources, three other overdense groupings were found in the diffuse area surrounding the massive cluster core.  However, these local overdensities had insufficient membership to allow any statistical conclusions, and subsequent analyses will focus on the main cluster core.  We report the position and membership count of all cluster cores in Table~\ref{cores}, and all cores are plotted in Figure~\ref{mst_grps}.

\section{DISCUSSION\label{discuss}}

\subsection{Nearest Neighbor Distances\label{nnd_disc}}

\begin{figure}
\epsscale{.80}
\plotone{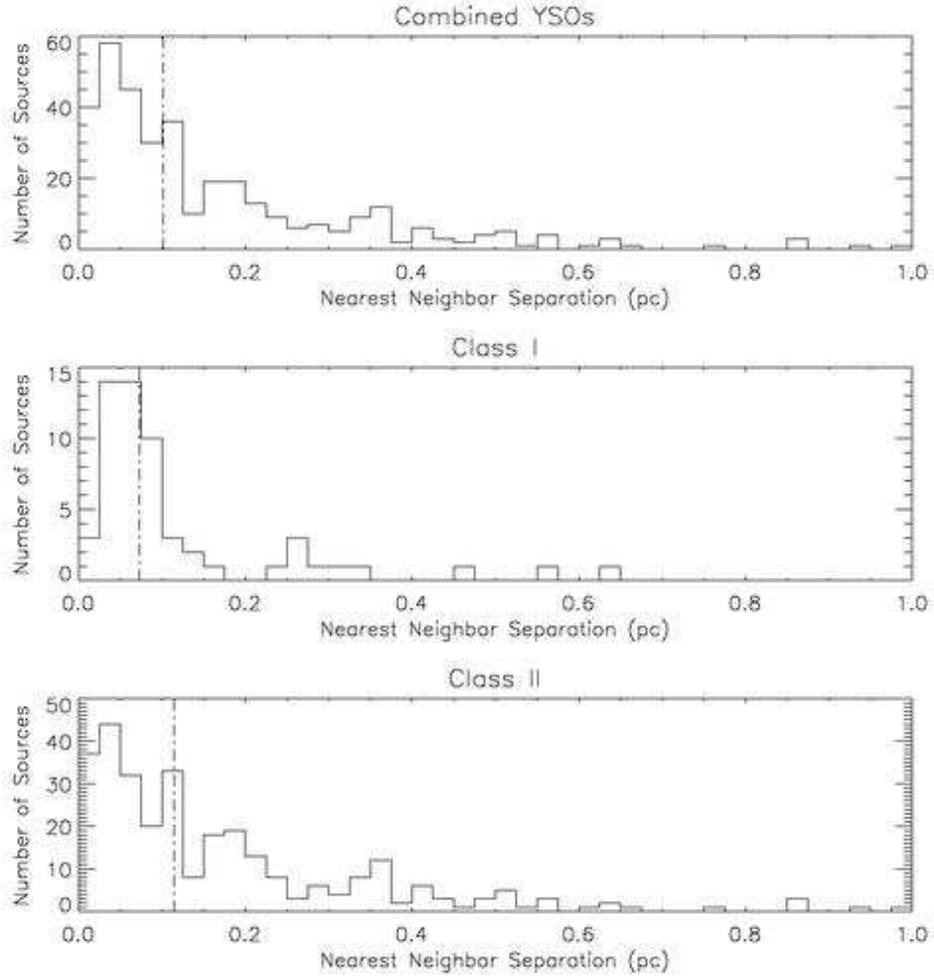}
\caption{Histograms of nearest neighbor YSO spacings, by class.
  Dot-dashed lines show the median nearest neighbor length in parsecs.\label{nnd_hist}}
\end{figure}

We present the distribution of YSO separations in Figure~\ref{nnd_hist}.  All separations are measured to the nearest YSO regardless of class, then are plotted separately by class as well as together.  There is a clear peak in the Class~I separation distribution at 0.05~pc; the Class~II distances peak at 0.035~pc, but have a much broader distribution than the Class~I separations.  This could be an effect of a subset of Class~II YSOs that are dynamically evolved and have moved away from their birth sites, partially erasing the primordial configuration we see in the well-defined Class~I distribution  peak.  The two-sided Kolmogorov-Smirnov null hypothesis probability for these data is $10^{-24}$, indicating that it is extraordinarily unlikely that the two distributions were drawn from the same parent distribution at random.

The nearest neighbor peaks for both classes, shown in Figure \ref{nnd_hist}, correspond closely to those found in similarly aged nearby clusters studied with {\it Spitzer}.  In NGC~1333 \citep{gut08}, all YSOs have nearest neighbor distances peaking between 0.02 and 0.05~pc, both separately by class and as a total YSO count.  While the peaks found in \citet{gut08} are similar to those found here, the Class~II spacing peak here is not prominent compared to the larger distribution.  In NGC~2264, nearest neighbor distances for Class~I YSOs peaked at a distance of 0.078~pc \citep{tei06}, but the study ignored Class~II YSOs that appear mixed among the protostars.

\subsection{Comparison with \citet{gut09}\label{b4} Cluster Survey}

With our expanded field of view, it is clear that the region is much larger than the area studied by \citet{gut09}.  We have also increased near-IR sensitivity in the densest region.  Previous coverage of the region was only large and deep enough to reveal 161 YSOs; this study has found 202 new YSOs (34 within the original field of view) and rejected three YSOs from the previous study, with a new total of 360.  Some YSOs changed class from those by \citet{gut09} considering the addition of new data.  Five objects previously considered deeply embedded protostars (2, 5, 7, 8, 9) are now more confidently classified as Class~I members, but in terms of the analysis by \citet{gut09}, this does not change the overall classification, as noted in Section~\ref{class}.  A small number of objects switched class from Class~I to Class~II (25) or vice versa (91, 130), since they lie close the color boundary between these classes.  For similar reasons, one object previously classified as Class~II (148) is now considered a transition disk YSO; two Class~II YSOs (51, 151) are now considered photospheric (ie. non-excess) sources; and one transition disk YSO (156) is now similarly classified as a photosphere.  Those exhibiting no excess emission are rejected as high confidence members, as they are most likely field stars, and are thus not included in the analysis presented below.  We have increased the Class~I count by 58\%, from 36 to 57, and found the region to contain 2.4 times as many Class~II YSOs, an increase from 125 to 303.  While this changes the region's Class~II/Class~I ratio from 3.5 to 5.3, the ratio within the cluster core increased only slightly, from 3.1 to 3.7.  The differences in YSO counts between \citet{gut09} and this study can be found in Table~\ref{regioncomp}.

\citet{gut09} compared physical size by measuring the effective radius, $R_{hull}$, defined as $\sqrt(A_{hull}~/~\pi)$, where A$_{hull}$ is the area of the core within the convex hull drawn in Section~\ref{coreext}.  With the expanded coverage, the region's effective radius is now 5.36~pc, 2.9 times larger than its previous radius of 1.84~pc.  While \citet{gut09} does not include large nearby clustered star-forming regions such as Orion or Cep~OB3b, the AFGL~490 region is considerably larger than the Mon~R2 region, the largest region in their study with 235 total YSOs and an effective radius of 2.02~pc.  It is worth noting that the full spatial extent of Mon~R2 as studied by \citet{gut09} may suffer from a truncation similar to that work's survey of the AFGL~490 region.  The full Mon R2 cloud, untruncated, is studied in \citet{gut11}.

Where the previous study found two small cluster cores in the AFGL~490 region with 130 YSOs combined, this study finds 219 YSOs in one large cluster core.  This is partly due to the increased field of view, enabling a complete survey of the zone of overdensity, and to the change in $L_{crit}$.  We added 80 YSOs to the higher-density area and 110 YSOs to the transition and background YSOs in the CDF.  This created a more shallow line fit in the background area of the CDF, causing the two fit lines to intersect at a longer length. As a result, $L_{crit}$ increased from 0.241~pc in \citet{gut09} to 0.319~pc, which caused the two cores found in \citet{gut09} to combine into one large core.  If we apply the $L_{crit}$ used by \citet{gut09}, we retrieve the two smaller cluster cores from that work; however, those cores are now spatially complete, and the extra sensitivity from SQIID fills in the area.  The new populations are 183 and 21 YSOs in \citet{gut09} cluster cores 1 and 2, respectively.

A summary of the comparison of the AFGL~490 cluster core structural parameters measured here to the 25th, median, and 75th percentile values for the entire cluster survey by \citet{gut09} can be found in Table~\ref{corecomp}. AFGL~490's cluster core is comparatively very large.  It contains 8.4 times more YSOs than the median YSO count from the survey (26), with 7.8 times as many Class~I members and 8.1 times as many Class~II members (6 and 21 being the median values from \citet{gut09}, respectively).  These values are outliers of those recorded by \citet{gut09}; however, the core Class~II/Class~I ratio of 3.7 is identical to the survey median.  The core not only has a large population but also an abnormally large spatial extent.  It has an effective radius of 1.68~pc, 4.3 times larger than the median radius (0.39~pc) and far larger than the 75th percentile of the survey cores.  Its aspect ratio of 2.07 is greater than the median aspect ratio of 1.82, making it more elongated than the typical core.
Despite its large physical size, the cluster core has a relatively low surface density of stars.  With a mean surface density of 22~sources~pc$^{-2}$, it falls even lower than the 25th percentile of the survey cores and is only 37.4\% of the median survey value of 60~sources~pc$^{-2}$.  The mean $A_{K}$ of 0.87~magnitudes is only slightly greater than the survey average of 0.8 magnitudes. 

\subsection{Correlation between Star Formation and the Natal Cloud\label{cloud}}

This cluster core resembles a normal core in all aspects but surface density and size when compared with the sample studied by \citet{gut09}.  The curious spatial distribution of YSOs in the AFGL~490 region led us to wonder whether the parsec-scale surface densities of YSOs correlate with associated gas column densities, as was observed in several other nearby molecular clouds \citep{gut11}.  In order to investigate this, we replicate their analysis using our YSO surface densities sampled at an equivalent spatial resolution as the extinction map, which we use to estimate the molecular gas column densities (Figure~\ref{av_map}).  We calculate YSO surface density by computing the 11th nearest neighbor surface density and sampling at each source's position; assuming an average of 0.5~M$_{\sun}$ per YSO, we can easily find the mass surface density in M$_{\sun}$~pc$^{-2}$.  To find gas column density, we first sample the nearest pixel in the extinction map to a YSO and use that value.  To get from extinction to gas column density, we assume that dust traces molecular gas and adopt the canonical conversion assuming N(H$_{2}$)/A$_{V}$~=~10$^{21}$~molecules~cm$^{-2}$~mag$^{-1}$ \citep{boh78}; thus, 1~A$_{V}$ corresponds to 15~M$_{\sun}$~pc$^{-2}$.  In this plot, the expected rate of contamination of 7 AGN per square degree would normally be used to limit the applicability of this plot, but because of the distance to the region, the contamination rate corresponds to a surface density too low to plot (0.03 contaminants~pc$^{-2}$).  An extinction limit is placed on the plot in order to reject data where $A_{V}$ is less than 1, where our signal to noise ratio is too low to rely on those extinction measurements.

The AFGL~490 region generally follows the power law behavior put forward by \citet{gut11}, confirming that the star formation in this region is driven by the structure of the cloud.  The low surface density of the YSOs therefore corresponds to a relatively low density of natal gas and dust.  Also, there is a decline in the Class~II/Class~I ratio as gas column density increases, meaning there are more Class~I stars in the denser part of the cloud.  We confirm this by performing a two-sided Kolmogorov-Smirnov test on the gas surface density values of the Class~II and Class~I samples; the resulting null hypothesis probability is $6.3 \times 10^{-4}$.  Similar trends were found in several other clouds studied by \citet{gut11}.  In summary, it seems that the anomalous cluster core size and YSO density compared to those of the other cores studied by \citet{gut09} is likely caused by atypical attributes of the cloud: the natal gas and dust is disorganized, dispersed over a large area with an intermediate column density rather than over a small area with high column density.

\subsection{Mass Segregation\label{offset}}
The intermediate-mass source AFGL~490 is visually offset from the center of the cluster core (see Figure~\ref{mst_grps}).  This raises questions about the primordial mass segregation of the cluster core, namely, does a core's most massive star form in the center, or does it migrate into the center after undergoing dynamical interactions?  A more detailed analysis may provide an interesting constraint on the degree of primordial mass segregation in this region.  \citet{kir11} recently investigated mass segregation in four nearby star-forming regions, finding that in most identified cluster cores the most massive star tends to be centrally located.  They also quantified the range of statistical variation expected for uniformly random positioning of the most massive member, that is, one that has no preferred position in its core.

Regions studied by \citet{kir11} are close enough to consist mostly of spectroscopically confirmed members, including the diskless sources that we cannot identify in this survey, as noted in data section above.  Regardless, the low Class~II to Class~I ratio in the AFGL~490 region suggests that this core is relatively young and has a relatively high disk fraction.  This minimizes the bias from using only IR-excess source member identifications \citep{her07}, making this study comparable to that of \citet{kir11}.  Fortunately, \citet{kir11} use the same method of cluster core extraction as in this paper, and so the groupings are comparably derived.

\begin{figure}
\epsscale{.80}
\plotone{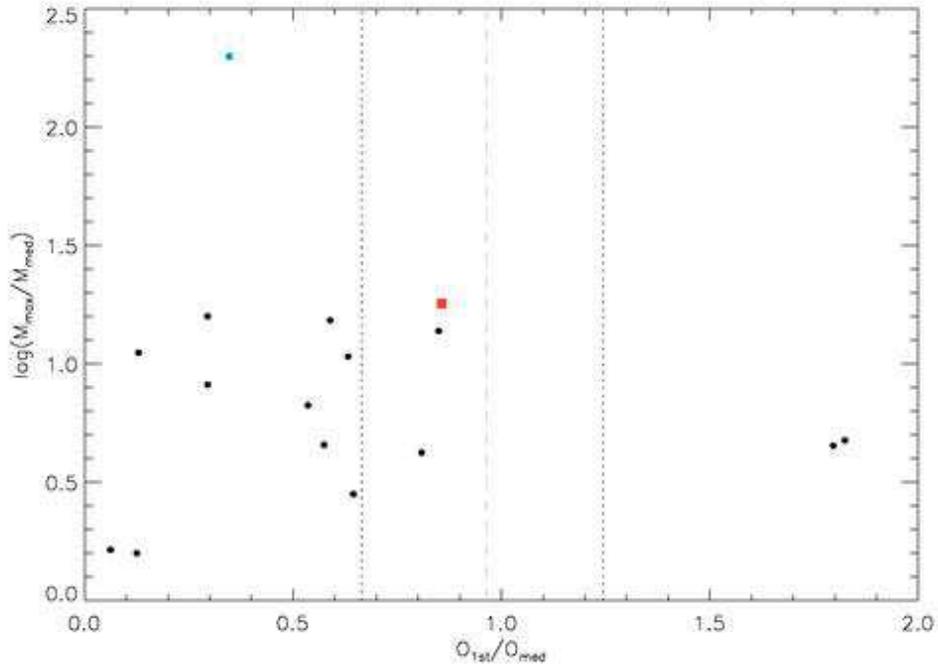}
\caption{Reproduction of Figure 9 of \citet{kir11}, a sample of four
  clusters whose cluster cores were determined in the same manner as
  this paper (black dots), with the AFGL 490 cluster core data overplotted (red
  square).  The Trapezium cluster in Orion, an example of a cluster
  with high mass contrast and centrally-located massive members, is
  plotted for comparison and can be seen as the outlier to the top
  (blue dot); Taurus and ChaI, whose most massive members are far from
  their median center, are the outliers to the right.  The x-axis plots the ratio of the radial offset of the most massive star to the median of the radial offsets of the known members in that cluster core; the y-axis plots the ratio of the mass of the most massive member to the median mass of the cluster core  members.  Dotted lines
  show 25th and 75th percentile values from simulations, while the
  dashed line shows 50th percentile \citep{kir11}.  Here, AFGL 490 is
  assumed to be 9~M$_{\sun}$ \citep{sch06}, while the median mass is
  assumed to be 0.5~M$_{\sun}$.\label{mass_seg}}
\end{figure}

\citet{kir11} measured the ratio of the offset of the most massive member of a cluster core to the median offset of the YSOs ($O_{1st}/O_{med}$) against the ratio of the mass of the most massive star in the core to the median mass of the YSOs in the core ($M_{max}/M_{med}$).  Using different methods to calculate the center of the cluster core, including mean and median positions of the members and the geometric center of the convex hull, we found the offset of the most massive source to be 0.74-0.95~pc.  The median offset of all YSOs from the center of the core was found to be 1.11~pc.  We adopted a mass of 9~M$_{\sun}$ for AFGL~490 \citep{sch06} and  0.5~M$_{\sun}$ as the median mass of the YSOs, assuming a normal initial mass function.  We plot the values for the fourteen cluster cores identified in \citet{kir11}, as well as the cluster core from this work, in Figure~\ref{mass_seg}.

We find that this cluster core's offset ratio falls between the 25th and 50th percentile of the simulated random position results found in \citet{kir11}, suggesting that we see no preferential position of the most massive member to the other members of the cluster core.  When we adopt the $L_{crit}$ used by \citet{gut09}, AFGL~490 still falls within Core 1; the offset ratio for this cluster core still falls within the expected statistical range for a non-preferential core position (with the data point moving right on the x-axis in Figure~\ref{mass_seg}), suggesting that this analysis is relatively secure regardless of the change in L$_{crit}$.

\section{SUMMARY\label{summary}}   

We have presented an analysis of the star-forming region AFGL~490, 900~pc distant.  It is an expansion of the work on a portion of this region reported by \citet{gut09}.  In summary:

\begin{itemize}

\item We have identified 360 YSOs in the entire star-forming region AFGL~490. Of those, 57 are identified as Class~I and 303 as Class~II YSOs.

\item We find that the distribution of nearest neighbor distances among these YSOs peaks in the same range as other similarly studied embedded clusters.

\item We isolated one denser subregion, or cluster core; it contains 219 YSOs, 60.8\% of the region's members.  The core is defined at a lower threshold surface density than that found by \citet{gut09}, effectively merging the two distinct cores previously reported in that work.

\item We have performed several structural measurements of the cluster core, finding it to be evolutionarily similar to most partially embedded cluster cores (Class~II/Class~I ratio of 3.7), 1.68~pc in radial size and elongated (aspect ratio of 2.07), with a surface density of 24.1~pc$^{-2}$, and partially embedded (mean A$_{K}$~=~0.87 magnitudes).

\item Of all the regions studied by \citet{gut09}, we find that this region is the largest in YSO count and effective radius, but also has a low surface density and no particularly dense area.

\item We find that despite the unusual size and surface density of the AFGL~490 cluster core, it bears the same YSO surface density to gas column density correlation shown by \citet{gut11} in eight nearby molecular clouds.  This result suggests that the large size and low surface density of the region are due to star formation in an unusually large, intermediate column density molecular cloud clump.

\item We find that the most massive member of the cluster core, AFGL~490, has no preferential position among the other members of the core according to the methods of \citet{kir11}.

\end{itemize}

\acknowledgments
This publication makes use of data products from the Two Micron All Sky Survey,
which is a joint project of the University of Massachusetts and the Infrared
Processing and Analysis Center/California Institute of Technology, funded by
the National Aeronautics and Space Administration and the National Science
Foundation.  This research has made use of the SIMBAD database, operated at CDS, Strasbourg, France.  This research has made use of the VizieR catalogue access tool, CDS, Strasbourg, France.  This work is based in part on observations made with the {\it Spitzer Space Telescope}, which is operated by the Jet Propulsion Laboratory, California Institute of Technology under a contract 1407 with NASA. Support for the IRAC instrument was provided by NASA through contract 960541 issued by JPL.  RAG gratefully acknowledges support from NASA grant NNX11AD14G.

{\it Facilities:} \facility{Spitzer}.

\begin{deluxetable}{cccccccccccccccc}
\tabletypesize{\scriptsize}
\setlength{\tabcolsep}{0.02in} 
\rotate
\tablecaption{{\it Spitzer}-identified YSOs: IRAC, MIPS, and Near-IR Magnitudes\label{sourcelist}}
\tablewidth{0pt}
\tablehead{\colhead{Region Name} & \colhead{Index} & \colhead{R.A.\tablenotemark{a}} & \colhead{Decl.\tablenotemark{a}} & \colhead{$J$} & \colhead{$H$} & \colhead{$K_S$} & \colhead{$[3.6]$} & \colhead{$[4.5]$} & \colhead{$[5.8]$} & \colhead{$[8.0]$} & \colhead{$[24]$} & \colhead{$A_K$\tablenotemark{b}} & \colhead{$\alpha_{IRAC}$\tablenotemark{c}} & \colhead{Class(G09)} & \colhead{Class} }
\startdata
AFGL490 & 1 & 03:27:15.48 & +58:50:00.6 & $17.80 \pm 0.04$ & $15.92 \pm 0.02$ & $14.74 \pm 0.01$ & $13.49 \pm 0.01$ & $12.92 \pm 0.01$ & $12.79 \pm 0.05$ & $12.47 \pm 0.09$ & $5.66 \pm 0.03$ & 1.07 & -1.74 & I* & I* \\
AFGL490 & 2 & 03:27:16.15 & +58:54:21.4 & ... $\pm$ ... & ... $\pm$ ... & ... $\pm$ ... & $12.86 \pm 0.01$ & $11.22 \pm 0.01$ & $10.48 \pm 0.01$ & $9.45 \pm 0.01$ & $5.78 \pm 0.03$ & ... & 0.93 & I* & I \\
AFGL490 & 3 & 03:27:17.19 & +58:50:32.8 & ... $\pm$ ... & ... $\pm$ ... & ... $\pm$ ... & $15.40 \pm 0.03$ & $14.02 \pm 0.02$ & $13.24 \pm 0.06$ & $12.13 \pm 0.10$ & $6.30 \pm 0.03$ & ... & 0.81 & I* & I* \\
AFGL490 & 4 & 03:27:22.61 & +58:43:59.7 & ... $\pm$ ... & ... $\pm$ ... & ... $\pm$ ... & $15.44 \pm 0.05$ & $14.24 \pm 0.04$ & $12.87 \pm 0.15$ & $12.31 \pm 0.16$ & $6.40 \pm 0.07$ & ... & 0.83 & I* & I* \\
AFGL490 & 5 & 03:27:23.59 & +58:55:07.9 & ... $\pm$ ... & ... $\pm$ ... & ... $\pm$ ... & $13.23 \pm 0.01$ & $12.21 \pm 0.01$ & $11.38 \pm 0.01$ & $10.71 \pm 0.04$ & $6.41 \pm 0.06$ & ... & 0.05 & I* & I \\
AFGL490 & 6 & 03:27:26.31 & +58:44:00.3 & ... $\pm$ ... & ... $\pm$ ... & ... $\pm$ ... & $16.47 \pm 0.08$ & $15.30 \pm 0.07$ & $14.58 \pm 0.19$ & $13.65 \pm 0.17$ & $6.95 \pm 0.08$ & ... & 0.32 & I* & I* \\
AFGL490 & 7 & 03:27:27.95 & +58:54:09.7 & ... $\pm$ ... & ... $\pm$ ... & ... $\pm$ ... & $11.83 \pm 0.01$ & $10.31 \pm 0.01$ & $8.99 \pm 0.01$ & $7.87 \pm 0.01$ & $1.82 \pm 0.01$ & ... & 1.70 & I* & I \\
AFGL490 & 8 & 03:27:29.66 & +58:44:20.3 & ... $\pm$ ... & ... $\pm$ ... & ... $\pm$ ... & $16.54 \pm 0.17$ & $13.37 \pm 0.03$ & $11.72 \pm 0.01$ & $10.38 \pm 0.01$ & $4.72 \pm 0.03$ & ... & 4.01 & I* & I \\
\enddata
\tablenotetext{a}{J2000 coordinates}
\tablenotetext{b}{Only provided for sources with valid $JHK_{S}$ or $HK[3.6][4.5]$ photometry.}
\tablenotetext{c}{Extinction is not accounted for in these values.  High extinction can bias $\alpha_{IRAC}$ to higher values.}
\tablenotetext{d}{Rejected as a member in this work.}
\end{deluxetable}

\begin{deluxetable}{ccccc}
\tabletypesize{\scriptsize}
\tablecaption{Median Position and Membership of Cluster Cores\label{cores}}
\tablewidth{0pt}
\tablehead{\colhead{Core} & \colhead{Members} & \colhead{Right~Ascension} &
  \colhead{Declination} & \colhead{Area (pc$^{2}$)}}
\startdata
1 & 219 & 51.819 & 58.747 & 8.917 \\
2 & 6 & 51.717 & 58.890 & 0.1017 \\
3 & 5 & 52.074 & 58.848 & 0.4991 \\
4 & 5 & 52.061 & 58.520 & 0.0818 \\
\enddata
\end{deluxetable}

\begin{deluxetable}{ccccccccccccccccccccc}
\tabletypesize{\scriptsize}
\rotate
\tablecaption{AFGL 490 Membership Statistics, Previous vs. Expanded Surveys\label{regioncomp}}
\tablewidth{0pt}
\tablehead{\colhead{Survey} & \colhead{$N_{cores}$} & \colhead{$L_{crit}$ (pc)} & \multicolumn{3}{c}{Total YSOs} & \colhead{} & \multicolumn{3}{c}{Class I} & \colhead{} & \multicolumn{3}{c}{Class II} & \colhead{} & \multicolumn{3}{c}{Class II / Class I} \\
\cline{4-6} \cline{8-10} \cline{12-14} \cline{16-18} \\
\colhead{} & \colhead{} & \colhead{} & \colhead{All} & \colhead{Core(s)} & \colhead{Bkgd\tablenotemark{a}} & \colhead{} & \colhead{All} & \colhead{Core(s)} & \colhead{Bkgd\tablenotemark{a}} & \colhead{} & \colhead{All} & \colhead{Core(s)} & \colhead{Bkgd\tablenotemark{a}} & \colhead{} & \colhead{All} & \colhead{Core(s)} & \colhead{Bkgd\tablenotemark{a}}}
\startdata
\citet{gut09} & 2 & 0.241 & 161 & 130 & 31 &   & 36 & 32 & 4 &   & 125 & 98 & 27 &   & 3.5 & 3.1 & 6.8 \\
Expanded & 1 & 0.319 & 360 & 219 & 141 &   & 57 & 47 & 10 &   & 303 & 172 & 131 &   & 5.3 & 3.7 & 13.1 \\
\enddata
\tablenotetext{a}{Background here refers to YSOs in the region not
  considered part of the cluster (not field stars), in keeping with Table 5 by \citet{gut09}}
\end{deluxetable}

\begin{deluxetable}{ccccc}
\tabletypesize{\scriptsize}
\tablecaption{Cluster Core Comparison, \citet{gut09} vs. AFGL 490\label{corecomp}}
\tablewidth{0pt}
\tablehead{\colhead{Property} & \colhead{25th Percentile} &
  \colhead{Median} & \colhead{75th Percentile} & \colhead{Expanded Survey}}
\startdata
No. Class I & 3 & 6 & 9 & 47 \\
No. Class II & 14 & 21 & 46 & 172 \\
No. YSOs & 20 & 26 & 55 & 219 \\
No. Class II/No. Class I & 2.0 & 3.7 & 8.0 & 3.7 \\
Effective Radius $R_{hull}$ (pc) & 0.26 & 0.39 & 0.56 & 1.68 \\
Aspect Ratio &  1.59 & 1.82 & 2.20 & 2.07 \\
Mean Surface Density (pc$^{-2}$) & 30 & 60 & 120 & 24.1 \\
Mean A$_{K}$ & 0.6 & 0.8 & 1.1 & 0.87 \\
\enddata
\end{deluxetable}

\begin{figure}
\epsscale{.80}
\plotone{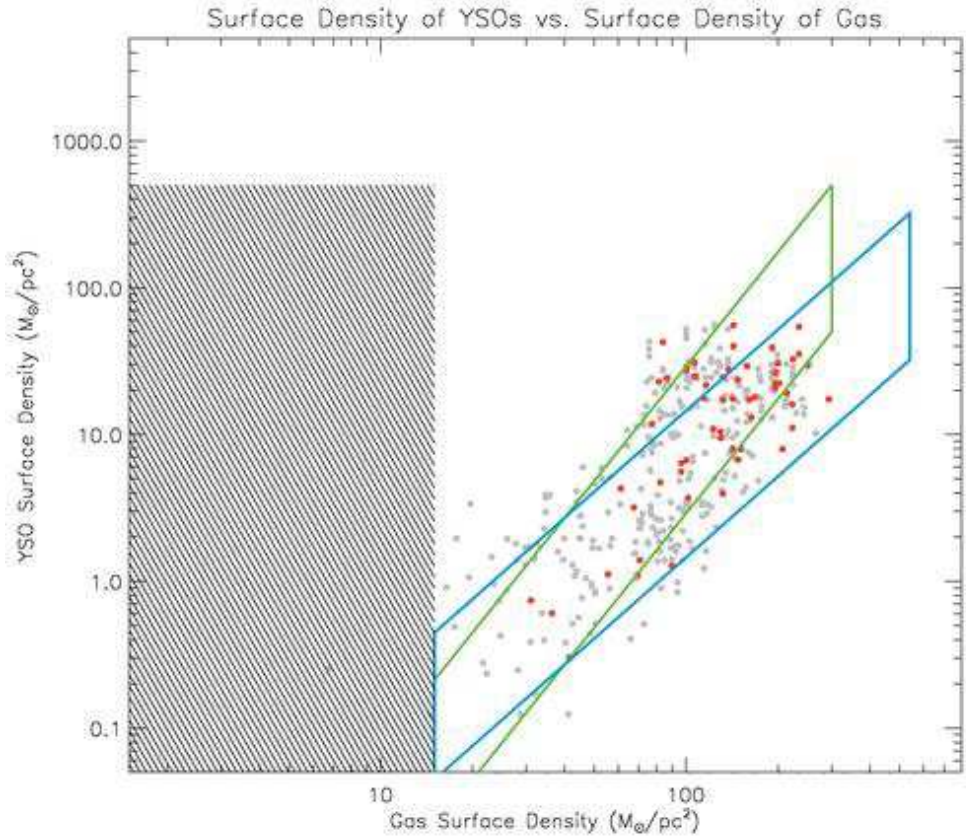}
\caption{ YSO surface density vs. gas column density, with Class I YSOs
  in red and Class II YSOs in grey.  YSO surface
  density is computed for each source by finding the NN11 surface density
  at the source's position.  Gas column density is measured from
  the extinction map at each source's position.  Areas overlaid with
  hashmarks are regions where $A_{V}$ is less than 1 and the angular
  surface density of sources is less than 7 per square degree (0.028, below the bottom axis limit).  The
  green and blue lines are fits to equivalent gas and YSO surface
  density measurements from surveys of the MonR2 and Ophiuchus
  molecular clouds, respectively \citep{gut11}.\label{nn11}}
\end{figure}

\end{document}